\documentclass[12pt,english]{article}

\usepackage[utf8]{inputenc}
\usepackage{graphicx}
\usepackage{setspace}
\usepackage{epstopdf}
\usepackage{subcaption}
\usepackage{authblk}
\usepackage[round]{natbib}
\usepackage{tabularx}
\usepackage[left=1.5in, right=1in, top=1in, bottom=1in, includefoot, headheight=13.6pt]{geometry}
\usepackage{amsthm}
\newcolumntype{b}{X}
\newcolumntype{s}{>{\hsize=.5\hsize}X}

\begin{document}

\begin{titlepage}

\title{Climate-related Agricultural Productivity Losses through a Poverty Lens} 
 
\author[1,2]{Alkis Blanz}

\affil[1]{\footnotesize{University of Leipzig, Faculty of Economics and Management Sciences, Grimmaische Str. 12, 04109, Leipzig, Germany}}
\affil[2]{Mercator Research Institute on Global Commons and Climate Change (MCC), Torgauer Str. 12-15, 10829, Berlin, Germany}

\date{\today}

\maketitle
\begin{abstract} 
In this paper, we analyze the long-term distributive impact of climate change through rising food prices. We use a standard incomplete markets model and account for non-linear Engel curves for food consumption. For the calibration of our model, we rely on household data from 92 developing countries, representing 4.5 billion people. The results indicate that the short-term and long-term distributive impact of climate change differs. Including general equilibrium effects change the welfare outcome especially for the poorest quintile. In the presence of idiosyncratic risk, higher food prices increase precautionary savings, which through general equilibrium affect labor income of all agents. Furthermore, this paper studies the impact on inequality for different allocations of productivity losses across sectors. When climate impacts affects total factor productivity in both sectors of the economy, climate impacts increase also wealth inequality.  
\end{abstract}

\thanks{\footnotesize{I am very grateful to Matthias Kalkuhl, Francesca Diluiso, Ottmar Edenhofer, Maik Heinemann and Ulrich Eydam for useful discussion and comments. Furthermore, I want to thank participants of the SURED 2022, the 27th Annual Conference of the European Association of Environmental and Resource Economists, the CESifo Area Conference on Energy and Climate Economics 2022 and the Potsdam Research Seminar in Economics for their useful feedback and comments.}}
\end{titlepage}

\restoregeometry

\newpage

\section{Introduction}

In recent decades, climate change has reduced global crop yields and slowed agricultural producticity growth \citep{OrtizBobea2021}. Relatedly, climate impacts already affect all dimensions of food security by disrupting production, quality, storage and retail \citep{shukla2019ipcc}. Agricultural productivity and food security are the most important resource for the livelihood of households in the developing world. Increasing weather and climatic events have exposed millions of people to acute food insecurity \citep{lee2023synthesis}. As developing countries will continue to be severely affected by climate impacts over a long time horizon, it is important to improve our understanding of the impact of agricultural productivity on long-term inequality.

The distributive impact of climate change in the short and long run may differ fundamentally. The direct, short-term impact treats income (and wealth) as fixed, and thus focuses on demand dynamics. In contrast, the indirect long-term impact considers changes in income and wealth through general equilibrium effects. Poverty is multidimensional, and assets play a detrimental role for the coping ability of households. In general, households with more assets tend to be more resilient and are able to cope better with shocks. In the presence of idiosyncratic risk, households accumulate assets in good times and deplete them in bad times to smooth consumption \citep{angus1991saving,rosenzweig1993credit}. For instance, wealth inequality could not only be affected by the direct impact on assets, but additionally may depend on how households will adjust their saving behavior in reaction to climate impacts. However, we still have limited knowledge about how climate change affects saving behavior across the wealth distribution and what the key driving forces are.

In this paper, we address this research gap by analyzing the long-term impact of food prices on inequality. By focusing both on the direct and indirect effects of climate-related food price changes, we aim to align micro- and macroeconomic analyses of the distributive impact of food prices in the developing world. In addition to the direct impact on consumption, higher food prices could affect the ability to accumulate assets over time. Food prices could increase consumption and wealth inequality. Within a single framework, we are thus able to compare the welfare impact of higher food prices in partial and general equilibrium. To properly model these dynamics, we extend the workhorse heterogeneous-agent macroeconomic model \citep{aiyagari1994uninsured,BEWLEY1977252,huggett1993risk} to account endogenously for nonlinear Engel curves for food consumption.  This allows to explicitly consider how both the consumption and the saving decision of heterogeneous households are affected by changes in food prices, while considering the impact of idiosyncratic risk. We calibrate the model to capture the characteristics of the developing world with 4.5 billion people, based on the Global Consumption Database from the World Bank.

 In the analysis, we proceed as follows. In a first step, we isolate the distributive impact of past agricultural productivity losses by comparing the developing world to a baseline, where without climate-related productivity losses, food prices are lower. There, we compare the short-term, partial equilibrium to long-term, general equilibrium outcome. While the direct impact focuses on changes in consumption, the general equilibrium outcome also considers changes in income. The poorest households suffer by far the largest direct impact. This is due to the fact that these households spend a large share of their expenditures on food. However, at the bottom of the distribution, we find the largest difference between partial and general equilibrium outcomes. Due to higher labor income, poor households can partially alleviate the large direct effect of higher food prices. In a second step, we compare the impact on inequality for different allocations of productivity losses across sectors. When productivity declines in both sectors, climate impacts also increase wealth inequality. 

\subsection{Related Literature}
 The key contribution is to bridge the gap between detailed micro-analyses of distributional effects of food price changes and general equilibrium model analyses of agricultural productivity losses. Thereby, this paper relates to several branches of the literature.
First, we contribute to the macroeconomic development literature has studied the nexus between food and agricultural productivity. In the developing world, a large part of the resources are used to produce food for subsistence needs \citep{gollin2007food}. Since food is a subsistence good, low agricultural productivity can limit growth, because high food prices adversely affect the savings rate \citep{irz2005seeds}. We contribute to this literature by studying how in the presence of idiosyncratic risk agricultural productivity losses affect the savings decision across the distribution.

By analyzing the impact of agricultural productivity losses through food prices in a heterogeneous agent macroeconomic model, this paper aims to bridge the gap to the microeconomic literature that analyzes the distributive impact of food price increases. Microeconomic methods and macroeconomic models generally find opposite effects of higher food prices on poverty \citep{headey2014food}. The general equilibrium effects are usually wage adjustments due to the agricultural supply-side response \citep{ivanic2014short,jacoby2016food}. Otherwise, households could adjust their labor supply to generate additional income to cope with higher food prices. For example,  households may increase their off-farm work to smooth consumption \citep{kochar1995explaining,kochar1999smoothing,heltberg2015}.

Furthermore, we relate to the literature that explicitly considers the fact that households in developing countries face a severe difficulty in insulating themselves from income volatility \citep{udry1994risk,townsend1994}. Existing evidence suggests that households may sell livestock or use savings when faced with health shocks \citep{ISLAM2012232,Isoto2017,deLoach2018} or when faced with droughts
\citep{FAFCHAMPS1998273,KAZIANGA2006413}. In developing countries, uninsurable idiosyncratic risk is relevant for welfare of households through its effect on risk sharing and migration \citep{lagakos2023welfare,morten2019temporary} and for the macroeconomy for agricultural productivity \citep{donovan2021equilibrium} and the severe limitations of risk sharing capabilities of privately owned businesses\citep{angeletos2007uninsured}. The contribution of this paper is to study the role of uninsurable idiosyncratic risk in the context of climate-related agricultural productivity losses. Thereby, we are able to study the impact on both consumption and wealth inequality.

 In the context of climate change, we build on the climate economic literature, where climate impacts are modeled as reductions in sectoral productivity (e.g. \citep{Schlenker15594}). This article studies agricultural productivity losses in a macroeconomic model with heterogeneous agents. Models with uninsurable idiosyncratic risk are increasingly used in the context of climate change \citep{Fried2021,malafry2022climate,Krusell2022}. By using a heterogeneous agent model, we are able to link this literature to the microeconomic literature that studies the distributive impact of increases in food prices related to climate change. \cite{hallegatte2017climate} combine household survey data from 92 countries with SSP scenarios and find that, through food prices alone, climate change could push 100 million people into extreme poverty by 2030. However, this widely used approach for studying the distributional incidence of climate or price shocks captures only the short-term, partial equilibrium effect and disregards how households accumulate assets and how labor income changes. Our general-equilibrium framework with heterogeneous agents allows households' consumption decisions to interact with macroeconomic dynamics. Thereby, we are able to to provide novel insights on how climate impacts affect the willingness and the ability to accumulate assets across the distribution.
 \\

The paper proceeds as follows.  Section 2 presents the dynamic incomplete markets model of food consumption. Section 3 discusses the calibration of the model. Section 4 reports the results of the quantitative results on the distributive impacts of rising food prices. Section 5 concludes.

\newpage

\section{A Dynamic Incomplete-Markets Model of Food Consumption}

Our model is at its core a standard incomplete market general equilibrium model with infinite horizon of the Aiyagari-Bewley-Hugget type \citep{aiyagari1994uninsured,BEWLEY1977252,huggett1993risk}. The special feature of the model studied here is that it accounts for Engel's law. As an endogenous outcome of the model, poorer households spend a larger share of their income on food. In the present model type, individual labor productivity is affected by an uninsurable idiosyncratic risk. This creates a precautionary saving motive and, through this, an endogenous wealth distribution. Higher food prices are an outcome of climate-related agricultural productivity losses. We abstract from aggregate uncertainty. Without aggregate uncertainty, in the baseline model agricultural productivity losses do not influence the idiosyncratic risk of households.

\subsection{Economic Environment}
\textbf{Households}\\ The economy is populated by a unit mass of infinitely lived agents whose preferences are represented by: 

\[
\sum_{t=0}^\infty \beta_t U(c_t,f_t)
\]
where
\begin{equation}
 U(c_t,f_t) = \frac{(c_{t}^\phi  (f_{t}-\underline{F})^{1-\phi})^{1-\eta}}{1-\eta}
 \label{util}
\end{equation}

Each agent maximizes utility (1) subject to the budget constraint
\[
\theta_{t} w_t + r_t a_{t} = c_{t} + p_t f_{t} + a'_{t}
\]

$f_t$ describes food consumption and $c_t$ non-food consumption of the household. $\eta$ is the coefficient of relative risk aversion (CRRA) and $\beta$ is the discount factor. Note that the price of food $p_t$ is expressed in units of non-food consumption. Agents can buy real, one-period bonds $a_t$ paying an interest rate $r_t$.  \\

The utility function has two properties that can be described as follows. The first part is a Cobb-Douglas structure between food and non-food consumption. This functional form implies a unitary substitution possibility between both goods. At the same time, this utility function is non-homothetic as agents have to satisfy at all times a subsistence level of food consumption $\underline{F}$. It can be thought of as the minimal amount of calories a human has to consume to avoid starvation. Thereby, the utility function has the well known Stone-Geary properties \citep{Stone1954,Geary1950} and are thus non-homothetic. Non-homothetic preference imply that the Engel’s law holds. Households spend a lower share on food, when households become richer. This property holds over time, when households get richer or poorer over time due to economic development. However, with heterogeneous agents Engels law holds also within each time period. Finally, both components may also interact. When poor households are close to the subsistence level, then substituting food consumption with non-food consumption becomes more difficult.


 This type of utility function implies nonlinear Engel curves and is widely used in the literature (for example \citep{kongsamut2001beyond,jacobs2019redistribution}), but not in the context of studying the distributive impacts of climate-related food price increases. There are multiple benefits from using this kind of utility function. First, the share of total consumption spend on food is endogenous in the model. Thereby, we can analyze how these shares change across the income distribution over time. Second, this kind of utility function is still analytically tractable. Finally, we can use micro data to estimate the parameters $\phi$ and $\underline{F}$ to replicate consumption pattern that are consistent with their empirical counterpart.

As in the standard incomplete markets model, agents face idiosyncratic risk to their labor productivity $\theta$ and experience transitory shocks to their labor endowment. As households supply one unit of labor inelastically, the time-varying labor productivity is the only source of difference in labor income $w_t \theta_t$ across households. As markets are assumed to be incomplete, this idiosyncratic labor risk is uninsurable. The income process itself is exogenous to the model. Together with the possibility of being borrowing constrained, this opens up a precautionary saving motive for households \citep{aiyagari1994uninsured}. The different history of labor productivity realizations yields an endogenous wealth distribution. When agents experienced a lot of low labor productivity realizations, their labor income is relatively low and thereby it is difficult for these households to accumulate assets. Note that this difficulty increase when the cost of living increase for households.

Apart from the idiosyncratic labor endowment risk, the possibility of being borrowing constrained is crucial for the precautionary savings motive. As in \cite{aiyagari1994uninsured}, we include a fixed borrowing limit in the form of $a'\geq \underline{a}$. We assume that the natural borrowing limit $\underline{a}^n$ holds. Even if an agent has the lowest labor productivity level $\theta_{min}$ at all periods (assuming that there is a positive probability that an agent with any current $\theta$ will keep receiving $\theta_1$), she can maintain a positive level of consumption. However, in the presence of the subsistence level of food consumption $\underline{F}$, we need to modify the natural borrowing limit. In our setup, it is not sufficient that non-food consumption is non-negative, but food consumption must satisfy the subsistence level.
    The budget constraint thus requires that:
    \[
    \underline{a}r +\theta_1 w = a' + \tilde{c} + p \underline{F}
    \]

    Once we solve for $\tilde{c}$ and assume it is slightly larger than 0. we get the adjusted natural borrowing limit:
    \[
    \underline{a}^n=\frac{w \theta_1 - p \underline{F}}{1-r}
    \]


\textbf{Production}\\

The production side of the economy is kept simple and consists of two symmetric sectors. The agricultural sector produces food and the other sector produces the non-food consumption good. As both sectors are symmetric production is described in both sectors by the following Cobb-Douglas technology:

\[
Y^j = A_j K_j^{\alpha_j} L_j^{1 - \alpha_j}
\]

where $j\in (c,f)$.\\

We assume that $\alpha_c=\alpha_f$ implies that the only difference between the sectors will be the level of total factor productivity. With levels of technology being the only difference, capital and labor are perfectly mobile across sectors. Thereby, we can focus on the demand side effects of rising food prices and isolate them from employment dynamics and structural change. For the same reason, we exclude the role of land ownership.\\

\textbf{Prices}\\

Intuitively, we assume that the non-food sector is relevant for investment. Considering the usual law of motion for capital, the setup implies that prices $w,r$ are given by:
\begin{equation}
r_t=1 +A_t^c f'(k_t)-\delta
\end{equation}

\begin{equation}
w_t=A_t^c[f(k_t)-f'(k_t)k_t] 
\end{equation}

 The useful feature of this setup is that food prices $p_t$ are entirely described by the Agricultural Productivity Gap $g_{apg}$, namely the differences in total factor productivity level between the two sectors:
\begin{equation}
p_t=\frac{A_t^c}{\tilde{A_t^f}}    
\label{eq:apg}
\end{equation}

where $\tilde{A^f}= (1-\xi^f) A^f$ is the agricultural productivity net of the climatic shock. In the baseline, climate change affects only agricultural productivity and $\xi^f$ describes the cumulative percentage loss due to athropogenic climate change. By lowering agricultural productivity $A^f$ , climate impacts widen the technological gap between both sectors and increase food prices. However, the parameter could be adjusted to account for trade responses or other adaptation measures. The idea is that $p$ are net food prices that are relevant for the consumer. In an extension, we consider additionally impacts on non-agricultural productivity $A^C$.

\subsection{Optimization and Steady State Equilibrium}

 The main decision of households is to choose total consumption expenditures over time. How households split up consumption between food and non-food consumption is static, meaning that it does not change over time. Instead of having two control variables $(c_f,f_t)$, we can simplify the optimization problem. It will be convenient to express the problem of households in indirect utility, as it simplifies the optimization problem. There the only control variable will be to choose income or total consumption expenditures, $y^{exp}=c + p F$. It will be useful for the analysis later on to distinguish between the decisions of households across and within periods. Using the definition of total expenditures, we derive the demand functions for non-food consumption and for food consumption:
\begin{equation}
    c_t = \phi (y^{exp}_t - p_t \underline{F})
    \label{demandC}
\end{equation}

\begin{equation}
    f_t = (1-\phi) \frac{y^{exp}_t}{p_t} + \phi \underline{F}
    \label{demandF}
\end{equation}

These demand functions describe how total consumption expenditures $y^{exp}$ are split up between food and non-food consumption within each period. Using these demand function the corresponding indirect utility function to \ref{util} is described by:
\begin{equation}
I(p,y) \equiv \frac{1}{1 - \eta}\bigg( \Phi  p^{\phi-1} (y^{exp} -p\underline{F})\bigg)^{1 - \eta}
\end{equation}

with  $ \Phi =\phi^\phi (1-\phi)^{1-\phi} $.

The optimization over the indirect utility function implies that the intertemporal decision is how to choose the total consumption expenditures $y^{exp}_t$, where as the intratemporal decision on how to split this consumption up into food and non-food consumption is governed by the demand functions derived above.\\

Given these prices the \textbf{Bellman equation} of households (in indirect utility) is:

\begin{equation}
V(a,\theta)= \max_{a'\geq 0} \frac{1}{1 - \eta} \bigg( \Phi  p^{\phi-1}\bigg)^{1 - \eta} (r a + \theta w - a' -p\underline{F})^{1 - \eta} + \beta E_{\theta'} V(a',\theta')
\end{equation}

subject to\\

$
a'\geq \underline{a}
$
$r$ and $w$ taken as given.\\

The Bellman equation balances the current reward of choosing a specific level of total consumption expenditures to the expected future reward. The future reward is affected by idiosyncratic risk. Depending on the future expected labor productivity $\theta'$, agents will choose incomes over time. The value function $V(a,\theta)$ considers the value a certain level of consumption will yield given that decisions are taken optimally. Solving the Bellman equations delivers policy functions. A policy functions describe optimal decisions rather than optimal sequences of consumption. In the present setting, policy function $y(a,\theta)$ describes the optimal level of savings given current realization of the labor productivity and current level of wealth.

Use the first order conditions and the Envelope condition to get:\footnote{A more detailed derivation can be found in appendix \ref{app_analytical}}
\[
 (y(a,\theta) -p\underline{F})^{ - \eta} = \beta  (y(a',\theta') -p\underline{F})^{ - \eta} r
\]
where 
\[
y(a,\theta)= r a + w \theta - a'
\]
and 
\[
y(a',\theta')= y(r a + w \theta - y(a,\theta),\theta')
\]

We start by studying the stationary equilibrium, where the wealth distribution is constant over time. However, even if the wealth distribution, and aggregates are constant, there is still movement on the individual level, where agents are hit by idiosyncratic shocks and adjust their savings behavior accordingly. For the numerical solution of the model, we rely on the endogenous grid method by \citet{Carroll2006}, which is widely used for solving models with heterogeneous agents.\\

Next, we turn to the description of the stationary equilibrium. A recursive competitive equilibrium consists of prices $r$,$p$ and $w$, the value function $V(a,\theta)$, the optimal decision rule $y(a,\theta)$, the wealth distribution $x(a,\theta)$, the aggregate capital stock $K$, and the aggregate labor supply $L$, such that:

\begin{enumerate}
    \item Households:
    Given prices $r$,$p$ and $w$, the value function $V(a,\theta)$ is a solution to the optimization problem of the consumer. $y(a,\theta)$ is the optimal decision rule.
    \item Firms:
    Prices $r$, $p$ and w satisfy the following conditions: $r = 1 + A^c f'(k) - \delta $ \\
    $w = A^c [(f(k)-f'(k))k]$\\
    $p = A^c/\tilde{A^f}$
    \item Consistency:
    $x(a,\theta)$ is a stationary distribution consistent with the optimal decision rule $y(a,\theta)$ and the associated Markov chain.
    \item Aggregation: 
    The aggregate capital stock is consistent with the stationary distribution $x(a,\theta)$ or:\\
    $K = \sum_{i=1}^{n_s} \int_A a \hspace{0.25cm} dx(a,\theta)$
\end{enumerate}

\newpage

\section{Calibration}

The model described in the previous section is calibrated to a large sample of developing countries. A strength of the model class we use is that many parameter values can be estimated based on micro data. However, due to data availability and quality issues in developing countries, we could not take all parameter values from the data. For the calibration of the consumption pattern, we relied on extensive household data. For the calibration of the main climate impact channel, namely through agricultural productivity, we rely on existing empirical estimates. The rest of the calibration follows the standard approaches in the macroeconomic literature.

The calibration is summarized in table 1.
\begin{table}
\caption{Parameter Values}
\begin{tabularx}{\textwidth}{bs}
\hline \hline
Parameter & Value \\
\hline
\textbf{Production}&  \\

Capital Share: $\alpha$ & 0.36 \\
Depreciation: $\delta$ & 0.08  \\
Agricultural Productivity Gap: $g_{apg}$ & 2.49 \\
\textbf{Preferences \& Consumption} & \\
Discount factor: $\beta$ & 0.975  \\
CRRA coefficient: $\eta$ & 2 \\
Persistence of Income shocks $\rho$ & 0.23\\
Subsistence Food level $\underline{F}$ & 0.0564\\
Food preference parameter: $\phi$ &  0.8196 \\
\textbf{Climate Impacts} &  \\
Baseline Productivity Loss & 0.25\\
Low Productivity Loss & 0.11 \\
High Productivity Loss & 0.4 \\
  \hline
\end{tabularx}
\end{table}

\subsection{Food Consumption and Preferences}
The first step is to use micro data to match the consumption pattern of households in the developing world. We rely on the Global Consumption Database \citep{GCD}, which is the most comprehensive data on household consumption expenditures covering 92 developing countries. These consumption pattern are reported based on four consumption segments\footnote{Four consumption segments are based on the following thresholds: Lowest$<\$2.97$ , $\$2.97<$  Low $<\$8.44$, $\$8.44<$ Middle $<\$23.03$, Higher $>\$23.03$ (per capita a day)}. The task at hand is to estimate the subsistence level of food consumption $\underline{F}$ and the preference parameter $\phi$. 
For the estimation of $\underline{F}$ and $\phi$, one can rearrange the demand function for non-food (\ref{demandC}) and food  consumption (\ref{demandF}) to get an expression for the food expenditure share:

\begin{equation}
    \frac{p*f}{y^{exp}}= \underbrace{(1-\phi)}_{\beta_0}  \hspace{0.25cm}  + \underbrace{p*\phi* \underline{F}}_{\beta_1} * \frac{1}{y^{exp}}
    \label{reg}
\end{equation}

Next, the model equation can be turned into a regression equation. 

Using the expenditure data from the Global Consumption Database, we regress the inverse of total consumption expenditures on the share spend on food in the data. As indicated in equation \ref{reg}, we then can calculate from the regression coefficients the desired parameter values. In detail, from the intercept we can calculate the preference $\phi$. Conditional on this value, one can then  calculate the subsistence level of food consumption $\underline{F}$. Notably, as the price of non-food consumption in the model is normalized to 1, a conversion factor is used to translate the subsistence level into model units. For this, we use the value of mean total consumption expenditures in the data. From the regression, we obtain $\phi=0.8196$ and $\underline{F}=0.0564$.

\subsection{Wealth and Income Inequality}
 The parameters of the income distribution are important for the endogenously determined stationary wealth distribution. As standard in the literature, we discretize the shock process and describe it as an N-state Markov chain. The stationary distribution of this Markov chain determines the income inequality in the steady state. Absent of aggregate uncertainty, the idiosyncratic risk to individual labor productivity is exogenous to the model and not affected by climate change. 

The idiosyncratic labor productivity risk can be estimated from the data. Where possible, both the standard deviation between labor productivity levels and the probabilities to change the productivity level can be obtained using household panel data. However, the panel data must have high quality and coverage over a long time horizon on the income sources of households. Such data are even rare in developed countries \citep{peralta2019welfare}. In the context of developing countries, the problem goes beyond the quality of available panel date. First, many households do not engage in employments with formal wage income. For example, subsistence farmers do not receive a wage each month that can be measured. Second, these arrangements are often in the informal sector. Thus, it is difficult to obtain the income dynamics of households in developing countries. Consequently, we abstract from estimating the idiosyncratic shock process directly from household data. 

In principle, we follow the calibration approach taken by \citet{malafry2022climate}. Inequality in the stationary equilibrium depends on a initial wealth endowment and the idiosyncratic labor productivity risk. Due to scarcity of high quality data on inequality for all countries of the sample, the calibration target is consumption inequality from the expenditure data used for the calibration of the preference parameter. In detail, inequality parameters are set to match the 80-20 ratio of total consumption expenditures from the Global Consumption Database. While in the data the 80-20 ratio of total consumption expenditures is $21$, the baseline steady state of the model produces 80-20 ratio of $20.94$. Thus, the model is able to match consumption inequality from the data well. The best model fit is achieved by using large differences in the spread of income shocks and low persistence. Robustness checks show that using more persistent income shocks with a lower spread does not affect qualitatively the main results of the analysis.

\subsection{Production and Agricultural Productivity}    
For the calibration, we are interested in the difference between the productivity levels in both sectors. The difference in technology levels determines the level of food prices in the baseline steady state. Following \cite{gollin2014agricultural}, we refer to this difference as the agricultural productivity gap. We use their database for the countries in our sample and calculate  the productivity gap for our sample and optain a value $g_{apg}=2.49$. This value corresponds to the agricultural productivity gap that accounts differences in hours worked and human capital.

The remaining parameters are chosen as standard in the literature. The capital share $\alpha$ is taken from \citet{peralta2019welfare} and is set to $0.37$. For the depreciation rate $\delta$, a range of values is applied within macroeconomic models that are calibrated to developing countries. We take an average of the values found in \citet{peralta2019welfare} and \citet{gollin2007food} set the depreciation rate to $0.08$. As the economy there is calibrated to a developing country, the parameter values chosen there seem to be suitable. However, the insights of our analysis are not sensitive on the calibration of these parameters.

\subsection{Climate Change}
Anthropogenic climate change has changed weather patterns globally in the past 50 years. These patterns are important for agricultural production. Most of the empirical literature focus on estimating the effect of climate change on yields of specific crops. However, due to our interest in sectoral dynamics, we are interested in the cumulative impact on the agricultural productivity. For the estimates of the cumulative impact of climate change on agricultural productivity in our sample, we rely on past estimated effects of climate impacts \citep{OrtizBobea2021}, which are reported in first columns of table 3.

The developing countries in our sample experience heterogeneous impact on agricultural productivity. Hotter countries, for example from Sub-Saharan Africa, tend to face large losses. In contrast, the losses in Central Europe or Asia could be significantly lower. In the baseline scenario, we focus on the aggregate productivity loss through climate change. First, we have to consider how many households in our sample live in which region of the world. This corresponds to the population-weighted average of the regional productivity losses. Thereby, we control for the regional differences and consider the average loss. 

 First, we consider their mean estimate as the baseline damage in our analysis. Furthermore, we consider as a robustness check the ($10\%$,$90\%$)-percentile as the optimistic and pessimistic estimate for climate-induced productivity losses. Thereby, we consider the uncertainty regarding the true loss in three different scenarios.With the baseline estimation agricultural total factor productivity is expected to drop by $25\%$. In contrast, the optimistic estimate yields a decline of only $11\%$, whereas the pessimistic estimate is $40\%$. The estimated loss in agricultural productivity is large. However, the final impact on productivity may additionally depend on regional substitution possibilities such as the location or the specific crop. The estimate is however broadly in line with future estimates net of adaptation. \citet{Hultgren2022} find that climate change, accounting for adaptation, will in regions lower agricultural productivity between 10 and 30 percent until 2100. Thereby, the present analysis may provide insights for future damages as well.
\begin{table}
\begin{tabularx}{\textwidth}{bssss}
\hline \hline
Region & Baseline & Optimistic & Pessimistic & Population Share\\
\hline
Africa & $-32.9$  & $-19.8$ & $-45.3$ & $28.2$ \\
Latin America \& Caribbean & $-30$ & $-19.4$ & $-44.1$ & $12.8$ \\
Near East \& North Africa & $-24.6$ & $-14.7$ & $-40.2$ & $4.7$ \\
Asia & $-21.4$  & $-12,4$ & $-37.4$ & $29.4$\\
Europe \& Central Asia & $-16$ & $5.2$ & $-33.4$ & $25.4$ \\ 
  \hline
  & & & &  \\
  
  weigthed Averages & $-11$ & $-25$ & $-40$ & \\ 
    \hline
   \hline
  \end{tabularx}
  \caption{Cumulative Impacts of Climate Change on Regional Agricultural Total Factor Productivity (in percent).}

\end{table}

\newpage

\section{Results}

The aim of this quantitative exercise is to analyze the nexus between climate-related sectoral productivity losses and long-run inequality. The starting point of the analysis is to focus on past agricultural productivity losses due to anthropogenic climate change. In the present analysis, we compare the actually observed inequality to a counterfactual world, where climate change did not affect agricultural productivity. In the counterfactual world, a higher agricultural productivity results in lower food prices. The first part of the quantitative exercise focuses on the distributive impact of changes in food prices in general equilibrium. The focus lies on decomposing consumption and income changes across the distribution. In a second step, we compare the results to the partial equilibrium outcome, where total consumption expenditures are held constant. The benefit of this exercise is that one can use closed-form solution from the model to compare the welfare impact with different levels of complexity. Thereby, one can isolate the impact of general equilibrium effects on welfare of poor and rich households.

The final subsection focuses on implications for future climate change. Climate change does not only affect agricultural productivity, but lowers simultaneously non-agricultural productivity. However, it is unclear how total productivity losses are distributed across sectors. The aim of this exercise is to analyze the distributional impact for different sectoral allocations of damages. An immediate consequence is that depending on the composition of damages different damage channels are active. These channels are directly linked to the agricultural productivity gap and differ in their impact on inequality.

\subsection{Distributive Impact in General Equilibrium}
This section aims to analyze the distributive impact of past agricultural productivity losses in general equilibrium. The present analysis relies on a comparison between two steady states. The difference between both steady states is the impact of anthropogenic climate change on agricultural productivity. We rely on estimates from \citet{OrtizBobea2021}, who calculate the cumulative impact of climate change on agricultural productivity from 1960 to 2016. The benchmark includes realized agricultural productivity losses. We compare this baseline to a counterfactual world, where climate change did not lower agricultural productivity in the developing world. 

In the world without climate damages, agricultural total factor productivity $A_f$ is higher is higher. As the productivity in the non-agricultural sector is unaffected, climate impacts affect the agricultural productivity gap. In detail, climate damages increase the agricultural productivity gap and thus food prices are higher (see equation (\ref{eq:apg})). Table \ref{tab:resBase} summarizes the baseline results of key aggregate variables to the decline in agricultural productivity due to climate change.\footnote{A comparison for the other low and high estimate can be found in appendix \ref{app_Comp}.} When we compare the world with climate change to a counterfactual world, where climate impacts did not affect agricultural productivity, we can observe the following. From a macroeconomic perspective, as a result of the agricultural productivity loss, output in both the agricultural and the non-agricultural sector decline. However, the decline in the agricultural sector is significantly larger. At the same time, aggregate capital is higher in the world with climate impacts.

The increase in aggregate capital is related to uninsurable idiosyncratic risk to labor productivity. The underlying savings mechanism in the standard incomplete market model are precautionary savings due to idiosyncratic labor productivity risk \citep{aiyagari1994uninsured}. In the presence of idiosyncratic risk, higher food prices affect additionally the precautionary savings motive. Apart from affecting the current ability to save, higher food prices increase the cost of living in periods, where households draw a low realization of the labor productivity shock. Relatively to the low income realization, the increase in cost of living is especially severe. Consequently, households want to increase the buffer savings for periods with low labor income. In settings with complete markets this channel is absent, because the individual labor productivity risk is insurable. With complete markets, higher food prices lead to lower aggregate capital accumulation. Changes in agricultural productivity only affect capital accumulation dynamics when preferences are non-homothetic.\footnote{See \citet{irz2005seeds} for a comprehensive analysis of the interaction between agricultural productivity and growth with complete markets.} In contrast, in the present analysis the steady state with climate damages features a higher capital stock due to the increase in precautionary savings.

 Before we turn to a more detailed analysis of the distributive impact, it is useful to analyze the underlying general equilibrium dynamics in more detail. The general-equilibrium dynamics do not depend including heterogeneous agents. The flow of reactions and adjustments are similar with a single representative agent (see for example \citet{irz2005seeds,wichmann2012agricultural}). Households react to the changes in food prices by adjusting both their food \ref{demandF} and their non-food consumption \ref{demandC}.\footnote{This direct impact of climate-related food price increases is studied in section 4.2 in more detail.}  However, apart from the partial equilibrium response, the changes in demand for food and non-food consumption affect production sectors. As demand for both goods change, also demand in the non-food sector change, which is used for investment. As labor is mobile across sector, this implies that the market-clearing wage rate and the interest rate change. In a second step, households adjust their saving decision in reaction these changes of factor prices. Consequently, steady states differ in their capital stock and output, because climate change permanently changed the productivity of agriculture. Note that these general equilibrium effects through changes in wages and interest rates are entirely driven by the initial change in \textit{demand} and not by climate change affecting wages and interest rates.\footnote{In a later exercise, we study the impact when climate change additionally affects non-agricultural total factor productivity.}

\begin{table}

\begin{tabularx}{\textwidth}{bsssssss}
\hline \hline
Scenario & $f^{8020}$ & $\mu^f$ & $y^{8020}$  &  $gini_W$ &  $\Delta Y^F$ &  $\Delta Y^C$ &  $\Delta K$\\
\hline
Baseline & $-1.11$  & $0.0311$ & $-3.13$ & $0.00008$ & $-0.1303$ & $-0.015$ & $0.028$ \\

\hline
    \hline
  \end{tabularx}

\caption{Baseline results for key aggregate variables: 80-20 ratio of food expenditures, mean share of food expenditures, 80-20 ratio of total consumption expenditures, Gini index wealth, agricultural output (percentage change), non-agriculture output (percentage change), aggregate capital (percentage change)}
 \label{tab:resBase}
\end{table}

Next, we turn to the results that are related to the distributional impact. At first glance, the distributive impact depends on the indicator. In a world with climate impacts, the difference in consumption expenditures is smaller between the richest and the poorest quintile, both for food expenditures and for total consumption expenditures. However, the decline is smaller for food expenditures, which is in line with the fact that food is a subsistence good. Intuitively, households spend on average a higher share of their expenditures on food,when food prices are higher. While inequality measured through total consumption expenditures decreases, wealth inequality, measured by the Gini Index, does not increase in response to agricultural productivity losses.

\begin{figure}
    \centering
    \includegraphics[scale =0.6]{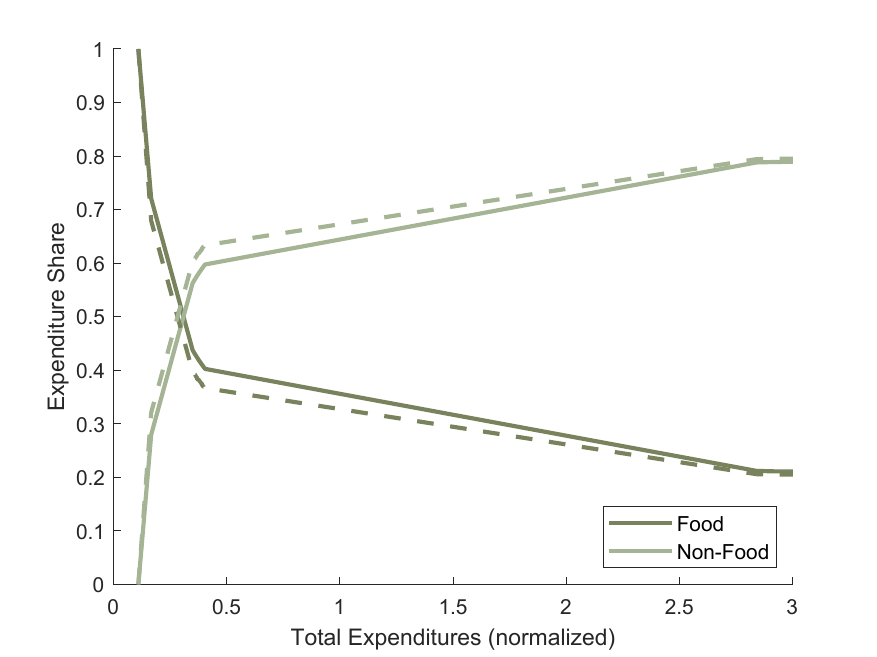}
    \caption{Consumption pattern across the distribution for steady state with damages (solid) and for steady state without damages (dashed), expenditures are normalized by mean level of  total consumption expenditures.}
\label{foodshareGE}
\end{figure}

As we are interested to understand how households \textit{across the distribution} adjust their consumption and savings decisions, our setting includes heterogeneous households. With heterogeneous households the aggregate demand dynamics differ compared to the representative agent case. Figure \ref{foodshareGE} summarizes the consumption pattern across the distribution for both steady states, thus including general equilibrium effects. In detail, the graph describes how both food and non-food consumption expenditure shares differ across the distribution. Total consumption expenditures are normalized by the mean level of total consumption expenditures. Thus, in the baseline steady state (solid lines), a household with total consumption expenditures equal to half of the mean level of total consumption expenditures spends 40 percent on food and 60 percent on non-food consumption. In contrast, in the world without climate impacts, the same household will spend only 36 percent on food due to lower food prices.\footnote{For the normalization of the new steady state, we use the mean of the baseline steady state. Thereby, the consumption pattern can be compared across steady states.} However, not all households adjust their consumption in a similar magnitude. In detail, households at the top and the very bottom
of the distribution adjust their consumption much less. Thus, intuitively, households spend a larger share of their expenditures on food, when climate impacts affect the agricultural sector.

\begin{figure}[h]
  \centering
  \begin{subfigure}{.5\textwidth}
    \centering
    \includegraphics[width=\linewidth]{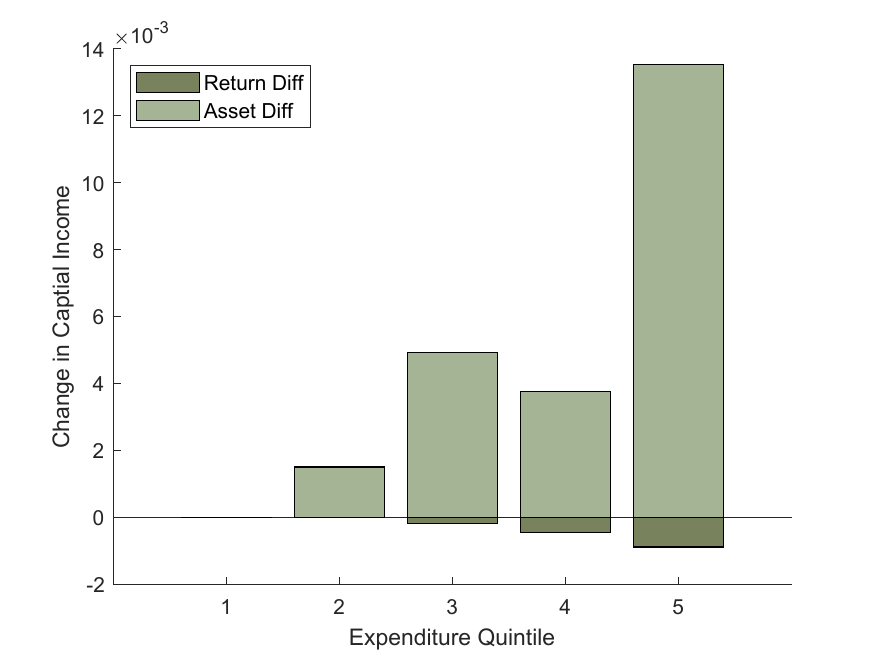}
    \label{decompCap}
  \end{subfigure}%
  \begin{subfigure}{.5\textwidth}
    \centering
    \includegraphics[width=\linewidth]{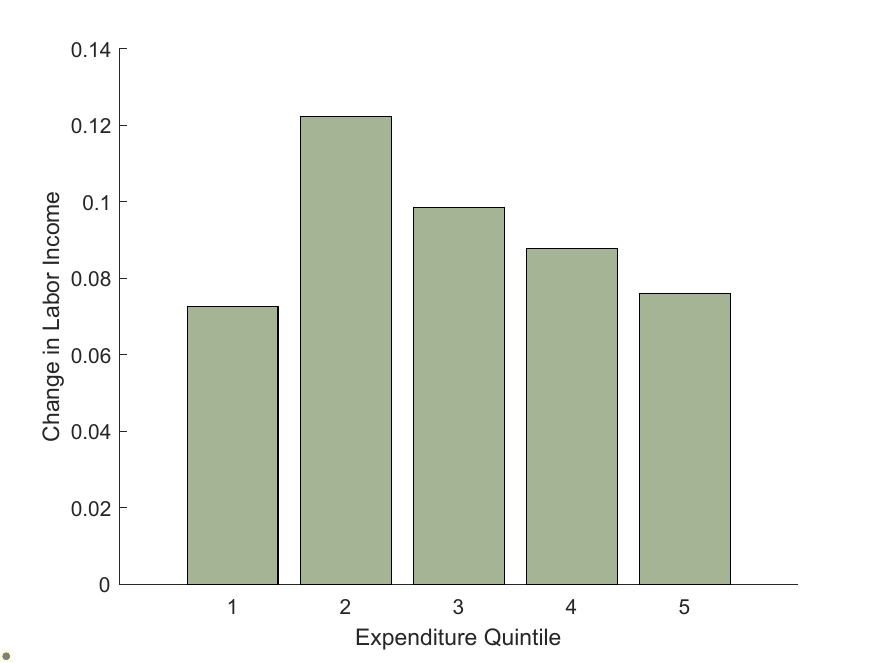}
    \label{decompW}
  \end{subfigure}
  \caption{Decomposition of income changes across the distribution for capital and labor income}
  \label{decompInc}
\end{figure}

This partial equilibrium effect differs due to non-linear Engel curves in food consumption. Households respond heterogeneously on the changes in food prices depending on their level of total consumption expenditures. Poorer households spend a larger share of their income on food and are thus more affected by price changes even if they face identical prices as richer households. Additionally, saving decisions will be affected heterogeneously across the distribution, as households differ in their initial wealth level, as well as in their dependency on capital and labor income. For understanding the long-term impact of food price changes it is not sufficient to study consumption expenditures. 

Essentially, the long-term distributive impact of higher food prices is driven by heterogeneous effects on incomes across the distribution. Depending on the direction of these changes in income, this could amplify or alleviate the short-term effect. For instance, if income in the steady state with climate damages is higher, households may increase their total consumption expenditures to cope with higher food prices. In general, there are two forces that affect savings behavior of households. First, when food prices are high the ability to save may be impacted due to non-homothetic preferences. Non-homothetic preferences imply that changes in food prices affect the cost of living of households directly. With higher costs of living it becomes more difficult for households to accumulate assets. Additionally, higher food prices affect the willingness to save in our setting. Thus, it is useful to analyze the income side of households in more detail.

In the steady state with damages, wages are slightly higher and interest rates are lower due to a higher capital stock. These changes in factor prices, will affect both labor and capital income. Figure \ref{decompInc} decomposes the income changes of households across the distribution relative to the world without climate change. For all households, total incomes are higher in the world with climate change. However, households across the distribution differ in their income sources. Intuitively, while richer households receive additional income from their asset holdings, poor households depend on labor income. Labor income $y_{inc,i}^w = w \theta^i$ is heterogeneous across the distribution. Even if the wage rate is the same for all households, as labor is assumed to be mobile across sectors, they differ in their individual labor productivity $\theta^i$ due to uninsurable idiosyncratic productivity risk. However, as we abstract from aggregate risk, the shock process for idiosyncratic risk is unaffected by climate change. Thus, we compare the same realizations of the labor productivity $\theta$ across steady states. Consequently, all households will benefit from the wage gain, but to a different degree based on the level of individual labor productivity.\footnote{Even in the steady state, individual labor productivity is fluctuating constantly. The present comparison is conditional on a specific realization of the shock.}

Capital income $y_{inc,i}^c =r a^i$ is also heterogeneous across the distribution, but on more dimensions. Even if the interest rate is common to all households, the capital holdings $a^i$ are not. In contrast to the realizations of the idiosyncratic labor productivity, households can adjust their capital holdings. We decompose the changes in capital income according to:

\[
\Delta y_{inc,i}^c = a_{NoDam}^i \Delta r + r_{Dam} \Delta a^i
\]

Both terms are specific to each household and together with the changes in labor income explain the differences across the distribution.
The first component $a_{NoDam}^i \Delta r$ describes the income change due to the changes in the return of assets. As the interest rate is lower in the new steady state, this term tracks the income loss for agents conditional on fixed initial capital holdings. This term increases with asset holdings. Thus, for poorer households the absolute income loss from the change in the return is small to zero, as they have very little or no asset holdings at all. The second component $r_{Dam}^{*} \Delta a_t^i$ describes the change in income that is associated with adjusting the asset holdings at the new return. As previously discussed, in the presence of idiosyncratic risk all households want to increase their savings. This term captures the the actual ability of households to increase their savings. The results indicate that in the baseline analysis most of the households are able to increase their asset holdings, albeit at different scale. In detail, for the richest quintile the change in asset holdings is the largest, while the poorest quintile still has zero asset holdings. In the middle of the distribution, households are able to increase their asset holdings as well.

When we study the demand-induced general equilibrium effects of climate-related food price changes, most households are able to increase their savings. The underlying driver are increases in labor income. Thus, poorer households benefit from higher wage income, because in the presence of idiosyncratic risk capital accumulation becomes more attractive. The aim of the present exercise is not to imply that climate impacts increase wages of the poorest households, but to isolate the impact of demand changes in the presence of incomplete markets and idiosyncratic risk. Depending on the strength of the precautionary savings motive and the size of the price increase, wage and capital income may change differently depending on the position in the distribution. However, higher incomes do not relate to higher consumption and thus welfare. Higher consumption expenditures in a world with climate impacts and thus higher food prices seems intuitive. To understand how welfare is affected across the distribution, it is useful to study consumption dynamics in more detail. Therefore, we turn next to the welfare analysis including a comparison to the partial equilibrium outcome.

\subsection{Comparison to Partial Equilibrium Outcome}

The aim of this section is to compare the previously discussed to the short-term, direct impact of higher food prices. The benefits are twofold. First, we can isolate the strength of the general equilibrium channel. Furthermore, one can study welfare changes across the distribution for different metrics. For the evaluation of the direct impact of higher food prices on can use the can use the steady state without damages to evaluate the direct effect of higher food prices.

For the partial equilibrium outcome the total consumption expenditures of each household are kept constant. Given the new prices, households can re-evaluate how they will split it up between food and non-food consumption. As we keep the total consumption expenditures for each household constant, the underlying model allows for deriving these dynamics analytically. Recall that the demand functions for food $f$ and non-food consumption $c$ are given by:\\

\[
    c_t = \phi (y^{exp}_t - p_t \underline{F})
\]
\[
    f_t = (1-\phi) \frac{y^{exp}_t}{p_t} + \phi \underline{F}
\]

Taking the partial derivative of the demand functions for food and non-food consumption with respect to food prices yields:\\

\[
\frac{\partial c(p)}{\partial p}= - \phi \underline{F}
\]
\[
\frac{\partial f(p)}{\partial p}= -(1 - \phi) \frac{y}{p^2} 
\]\\

For increasing food prices both derivatives are strictly negative. Thus, the immediate response of households to higher food prices is to consume less of both food and non-food goods. While the decrease of non-food consumption is the same across the expenditure deciles, the decline in food consumption is not. The decline increases with the level of total consumption expenditures. Richer households reduce their food consumption more strongly than poor households. Consequently, households end up adjusting differently their share of total consumption expenditures spent on food:\\

\[
\frac{\partial f(p)p/y}{\partial p}= \frac{\underline{F} \phi}{y}
\]
\\

This emphasizes the role of non-homothetic preferences. Poorer households spend a larger share of their total consumption expenditures on food and are closer to the subsistence level. The direct impact on consumption of food and non-food consumption is larger than the observed changes including general equilibrium effects. Notably, the difference between the partial and general equilibrium analysis is largest for the poorest households. While this is an indication that changes in total consumption expenditures matter for these households, the consumption pattern cannot be translated directly to welfare. Thus, we next compare changes in welfare across the distribution for the partial and the general equilibrium outcome. \\

\begin{figure}
    \centering
    \includegraphics[scale=0.7]{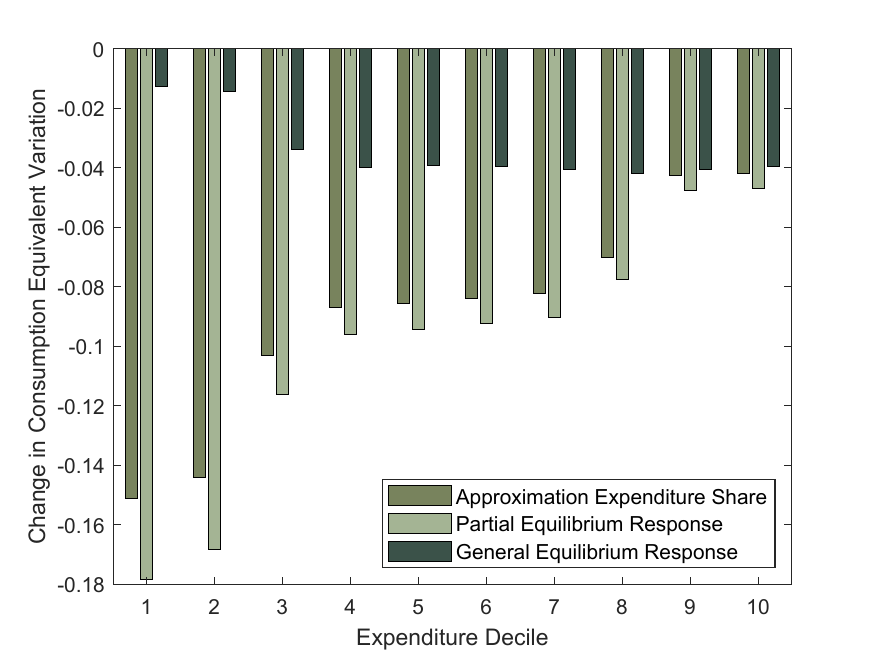}
    \caption{Change in consumption equivalent variation for the partial and general equilibrium outcome and for an approximation based on the initial food expenditure share.}
\label{welfare}
\end{figure}

Figure \ref{welfare} shows the welfare outcome in partial equilibrium in terms of consumption equivalent variation.\footnote{The formulas are included in appendix \ref{app_analytical}.} In partial equilibrium, poor households suffer the most from lower food prices, as they are closer to the subsistence level of food consumption and initially spend a larger share on food. Households at the bottom of the distribution are sensible to changes in wages due to their dependence on labor income. These households do not own any wealth, they are insensitive to changes in the interest rate. For the poorest households, the large real income decrease from the direct effect is dampened by an increase in labor income. The higher labor income is due to a higher wage rate in the new steady state. Even if the increase in the wage rate is small, it has an impact on the poorest households given their very low total income level. While poor households can consume less for a given level of consumption expenditures (partial equilibrium effect), they are partly able to cope with this by increasing their total consumption expenditures due to higher wage income.

Similar to the poor, the welfare impact for households in the middle of the distribution differs in partial and the general equilibrium. The difference is driven by changes in total consumption expenditures. Households in the middle of the distribution increase their consumption expenditures to cope with higher food prices. While these households benefit too from higher wage income, these households additionally receive capital income. In contrast to the poorest households, households in the middle of the distribution are able to increase their capital holdings. Thus, the difference in welfare between partial and general equilibrium for these households is lower, as they increase not only their total consumption expenditures, but also their capital holdings due to a higher precautionary savings motive.

For the richest households, the difference in welfare between partial and general equilibrium is negligible. First, the direct impact of higher food prices on these households is small. In detail, as rich households spend a smaller share on food, they are less affected by higher food prices. Furthermore, the income of the richest households increase compared to the baseline, but relatively less then for poorer households. Due to the stronger precautionary savings motive and the smaller direct impact, these households prefer to increase their capital holdings rather than their consumption expenditures. Consequently, total consumption expenditures and thus the welfare impact for these households does not differ significantly between the partial and general equilibrium outcome.

We compare it to an approximation based on the initial consumption expenditure share spend on food. The idea is to compare the partial equilibrium effect to a reduced form approximation. The approximation is based on the initial food expenditure share and the percentage change in food prices. Thus, the the approximation relies simply on Engels law and does not use the explicit demand functions. The results indicate that Engels law is a suitable approximation for the short-term welfare impact of climate change on inequality through changes in food prices. However, the initial share spent on food is not a suitable predictor for how households will adjust their total consumption expenditures when food prices increase. As the adjustment of total consumption expenditures relies on heterogeneous income changes across the distribution, the approximation is unable to capture the general equilibrium welfare impact.

\begin{table}[ht]
\begin{tabularx}{\textwidth}{bsss}
\hline \hline
Expenditure Decile & Low Estimate & Baseline & High Estimate  \\
\hline
1 & $-0.071$  & $-0.136$ & $-0.189$ \\
2 & $-0.067$  & $-0.128$ & $-0.177$ \\
3 & $-0.039$  & $-0.072$ & $-0.098$ \\
4 & $-0.029$  & $-0.051$ & $-0.069$ \\
5 & $-0.0287$  & $-0.05$ & $-0.068$ \\
6 & $-0.027$  & $-0.048$ & $-0.063$ \\
7 & $-0.027$  & $-0.46$ & $-0.061$ \\
8 & $-0.019$  & $-0.033$ & $-0.043$ \\
9 & $-0.008$  & $-0.007$ & $-0.007$ \\
10 & $-0.008$  & $-0.007$ & $-0.006$ \\

\hline
\end{tabularx}
\caption{Comparison of change in welfare between partial and general equilibrium outcome for baseline productivity loss ($\xi^f=0.25$) for low ($\xi^f = 0.11$) and high estimates ($\xi^f = 0.4$). Change in welfare is calculated as the difference in consumption equivalent variation between partial and general equilibrium outcome.}
\label{tab_PEGE}
\end{table}

 Next, we perform a sensitivity analysis to the main result. The main focus of the sensitivity analysis is the impact of climate change on agricultural productivity. The estimation of the agricultural productivity losses include some uncertainty. The hotter a region is the more uncertainty surrounds the estimate of the productivity loss \citep{OrtizBobea2021}.
 As more developing countries in our sample are in hotter regions, which are associated with higher uncertainty in the estimate, we consider three different agricultural productivity losses. The baseline loss in the main analysis is the mean estimate, where as here we consider the ($10\%$,$90\%$)-percentile as the optimistic and pessimistic estimate for climate-induced productivity losses. We consider this to be the main point for sensitivity analysis for several reasons. First, it is the channel through which climate change enters the analysis and determines fully the change in food prices. Furthermore, the cumulative total factor productivity losses are significant in the baseline analysis. In this sensitivity analysis, we want to ensure that the results are not driven by the size of the decline. Finally, the main result builds on the general equilibrium effects and the shock size is a main determinant of their strength. As one can keep everything else constant between the steady states, we abstract from changes to structural parameters as then also the baseline steady state would change and we would be unable to do a comparison to the main results.

 Table \ref{tab_PEGE} reports the results of the sensitivity analysis.\footnote{A graphical comparison of the difference in welfare per decile between partial and general equilibrium outcomes for low and high estimate can be found in appendix \ref{app_fig}.} The main focus of the sensitivity analysis is the change in welfare. In detail, we compare the difference in consumption equivalent variation between the patrial and general equilibrium outcome for each expenditure decile. For example, if the difference is negative, it indicates that the partial equilibrium welfare loss is larger than the general equilibrium welfare loss. The results show that the partial equilibrium welfare loss is larger in all scenarios. Thus, in all scenarios, through a stronger precautionary savings motive, incomes of all households increase. Intuitively, general equilibrium effects and thus the impact on the incomes depend on the size of the productivity loss. A higher productivity loss implies a stronger precautionary savings motive and through higher savings o a higher wage income for all households. However, quantitatively, we find differences across the distribution.

 Across all sizes of productivity losses we can observe the strongest differences between partial and general equilibrium effects at the bottom of the distribution. Thus, the potential to overestimate the welfare loss of the poor based on the direct impact of food prices increases with the size of the productivity loss.  Through higher incomes the poorest households, who are not able to accumulate assets, the poor can adjust their total consumption expenditures to cope with higher food prices. This adjustment in expenditures drives the difference between partial and general equilibrium welfare for the poorest. In contrast, households at the middle of the distribution will partly also increase their asset holdings due to the stronger precautionary savings motive. As these households still spend a significant share of their income on food, they will increase their total consumption expenditures to cope with higher food prices. However, as these households are able to increase their savings, they will use part of their higher income to increase their asset holdings due to a stronger precautionary savings motive. Thereby, households at the middle of the distribution increase their buffer savings for periods with low labor income, where due to climate change the cost of living (through food prices) will be higher.

For the households at the top of the distribution, we do not observe differences across the shock sizes. For these households the difference between partial and general equilibrium analysis is similar across all estimates. This indicates that these households are always able to cope with the negative impact from higher food prices without adjusting their total consumption expenditures. The richest households simply increase their asset holdings due to a higher precautionary savings motive. Thus, for the richest households the partial and general equilibrium welfare outcome is identical.

\subsection{Climate Impacts and the Agricultural Productivity Gap}
In the previous section, we compared the distributive impact of climate-related food price changes in partial and general equilibrium. While the previous sections relied on estimates of the past impact of anthropogenic climate change on agricultural productivity losses, the underlying dynamics are also relevant for future climate change. In the following section, we want to explore how a potential bias towards agricultural productivity losses affect the impact channels of climate change. As the sectoral composition of damages is uncertain at best, we have to think what the implications of asymmetric impacts are.

\begin{figure}
    \centering
    \includegraphics[scale =0.5]{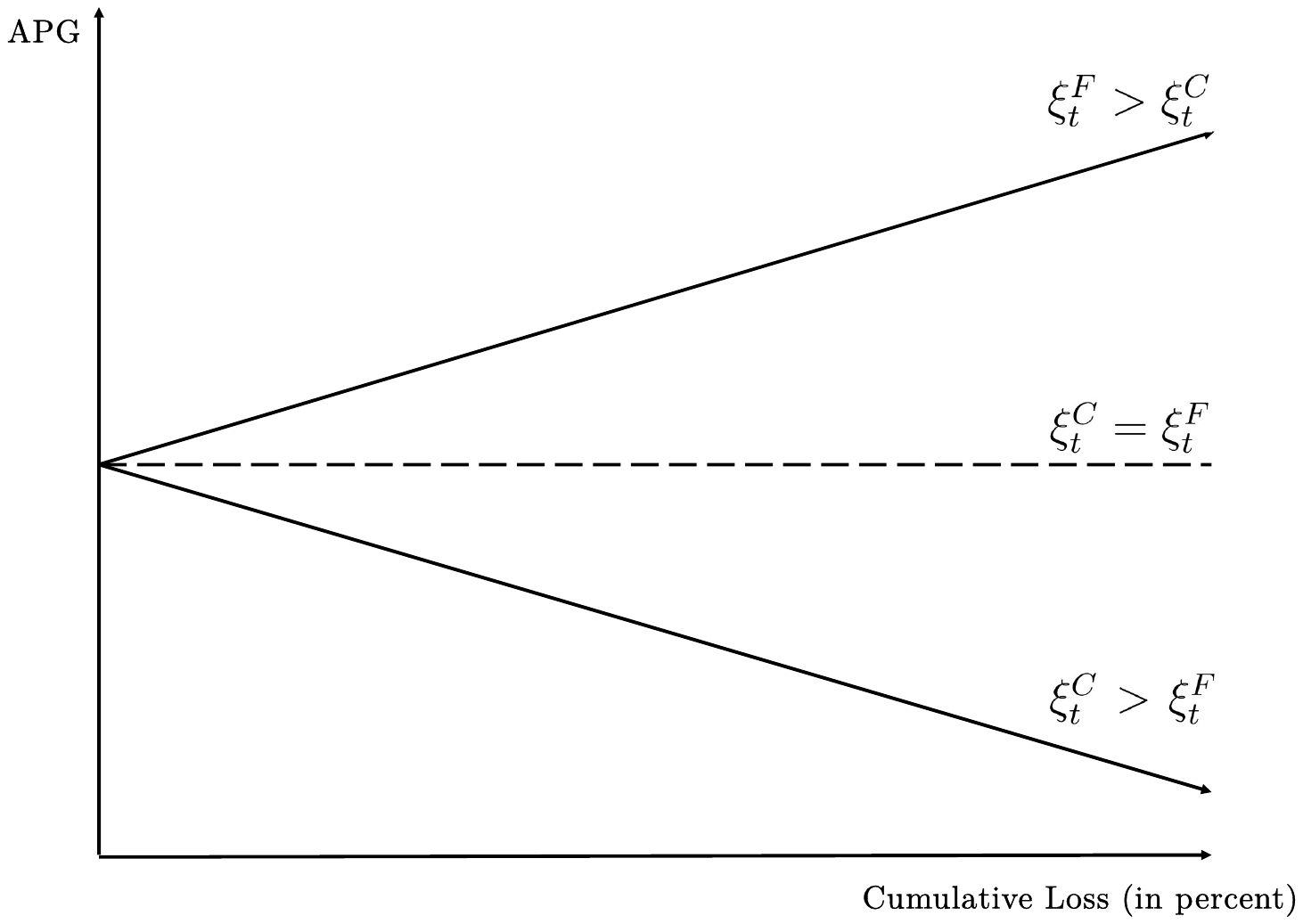}
    \caption{Changes Agricultural Productivity Gap (APG) depending on the distribution of damages across sectors}
\label{apg}
\end{figure}

In principle, climate change may lower  the level of total factor productivity in the agricultural sector $\tilde{A^f}= (1-\xi^f) A^f$ and in the non-agricultural sector $\tilde{A^f}= (1-\xi^f) A^f$. The parameters $\xi^f$ and $\xi^c$ capture the respective loss in total factor productivity due to climate change. However, it is uncertain if both sectors are affected equally or if climate impacts affect one sector more strongly. In our setting, the potential asymmetry in impacts is related directly to the agricultural productivity gap. The agricultural productivity gap describes the difference in productivity levels between agriculture and non-agricultural sectors. Figure \ref{apg} illustrates how different distributions of damages across sectors affect the agricultural productivity gap. Asymmetric climate impacts can widen or lower the agricultural productivity gap. As we will see this will imply different impact channels that can affect households heterogeneously.

The reference point is the case of symmetric productivity damages. With symmetric damages agricultural and non-agricultural productivity are affected equally, i.e. $\xi^c=\xi^f$, where $\xi^f, \xi^c > 0$. Consequently, the agricultural productivity gap stays constant. In this case, the damage channel is identical to a single sector model. There is no direct impact on households through changes in food prices. Households are affected through changes in factor prices. The strength of this impact depends on the size of the change in non-agricultural productivity due to climate change ($\xi^c$). As households depend heterogeneously on labor and capital income, the indirect impacts can affect inequality. Even if damages do not hit households directly inequality can increase in the long run.

Next, we turn to the case, where the agricultural sector bears a higher share of the productivity losses, i.e. $\xi^f > \xi^c$. In this case, there are two separate impact channels present. The first impact channel is the previously described impact through changes in factor prices. The second impact channel is related to the change in the agricultural productivity gap. As agricultural productivity falls more than non-agricultural productivity, the agricultural productivity gap widens. A higer agricultural productivity gap translates to higher food prices. Thus, apart from the indirect impacts through changes in factor prices, households are additionally affected by an increase in consumption prices. The impact through consumption prices will additionally affect factor prices. As seen in the previous section, where this channel was studied in isolation ($\xi^c=0$), demand changes and heterogeneous savings decision trigger a general equilibrium response. However, when non-agricultural productivity is lower due to climate change (but still less affected than agricultural total factor productivity), then the demand-driven general equilibrium effects interact with the climate-related impact on factor prices.

\begin{figure}
    \centering
    \includegraphics[scale=0.7]{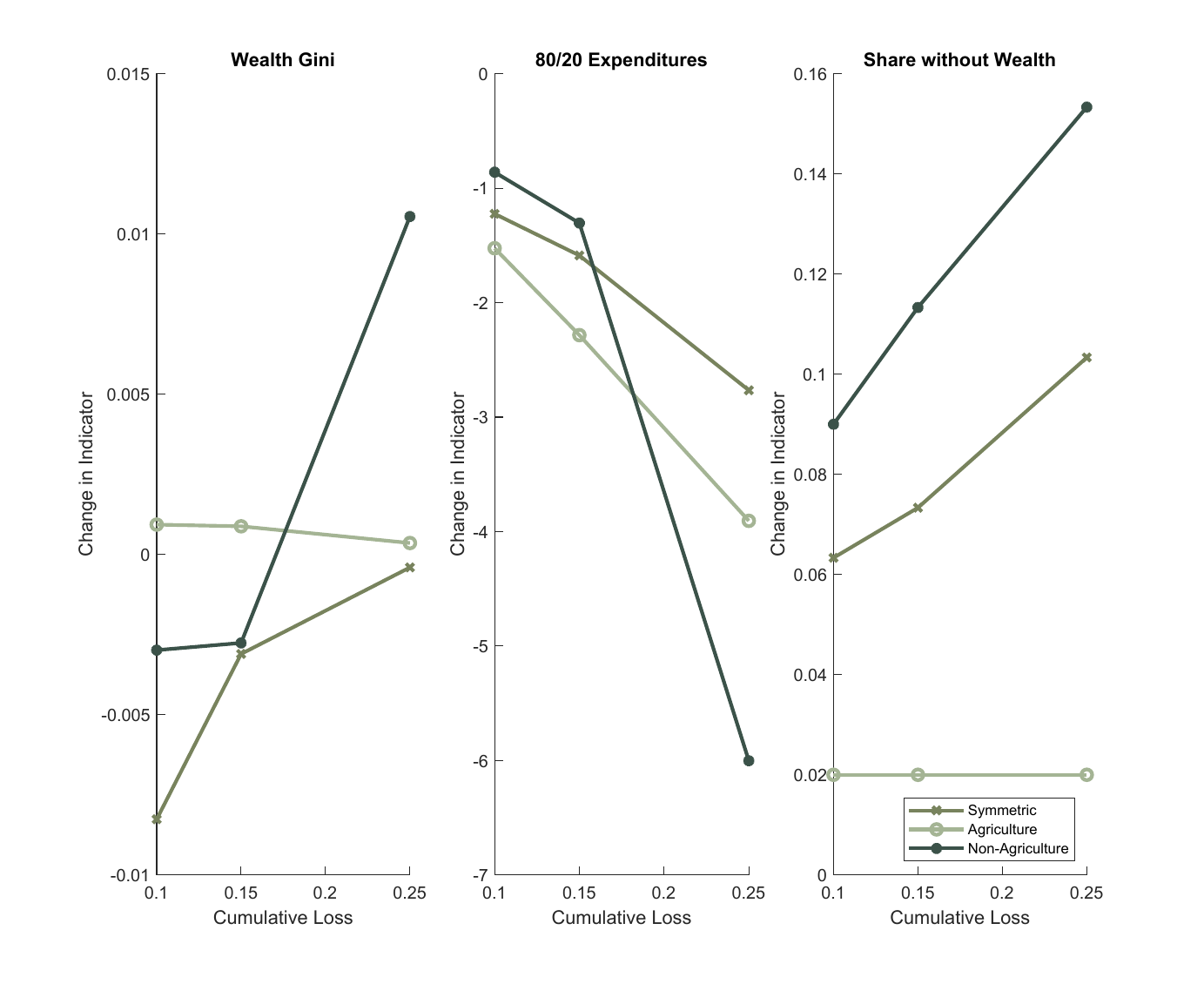}
    \caption{Changes in poverty indicators for different allocations of sectoral productivity losses.}
\label{indicators}
\end{figure}

In the last case, the non-agricultural total factor productivity is affected more by climate change, i.e.  $\xi^c > \xi^f$. While in principle the same impact channels are active, some interesting dynamics emerge. A higher impact on non-agricultural productivity implies a lower agricultural productivity gap. A lower agricultural productivity gap translates in our setup to lower food prices. Thus, even if there is a strong impact on the income side of households through changes in factor prices, lower consumption prices may alleviate some of the impacts on the expenditure side.

In the following, we analyze the impact of each of these channels in isolation. Thereby, we aim to provide a basic understanding of their potential distributional impact. The reference point is the case of symmetric productivity losses. We compare this baseline to the extreme cases. First, we consider the case of the previous section, where damages only affect agricultural productivity ($\xi^c=0$). In this case, we can study in isolation the general equilibrium impact of demand side dynamics through changes in consumption prices. Second, we consider the case where only non-agricultural productivity is affected ($\xi^f=0$). In this case, households benefit on the expenditure side from lower food prices, but face instead lower incomes through climate impacts. Finally, in the case of symmetric damages, the consumption expenditure damage channel is muted and households only are affected through direct impacts on incomes.

Figure \ref{indicators} summarizes the impact on poverty indicators for each of the scenarios. In detail, the comparison includes cumulative productivity losses from 10 to 25 percent for each of the scenarios. Productivity losses in the non-agricultural sector are weighted by the initial output share of non-food output. Thereby, the losses in agricultural and non-agricultural productivity are comparable. In the present analysis, we rely on the wealth Gini index, an 80-20 ratio for total consumption expenditures, as well as the share of households that have no wealth. The change in inequality in total consumption expenditures captures the changes in consumption inequality, which is relevant for welfare. In contrast, through the combination of the share of wealthless households and the wealth Gini index allows for a more elaborate focus on wealth inequality.

When climate damages hit only the agricultural productivity, the dynamics are similar to the previous exercise. Higher food prices increase the precautionary savings motive. However, not all households have the ability to increase their savings. The poorest households, who initially do not hold any assets, are not able to increase their asset holdings. Instead, these households increase their consumption expenditures to cope with higher food prices. As richer households are not affected that strongly by higher food prices, richer households increase their asset holdings rather than their consumption expenditures. Consequently, inequality measured by the 80-20 ratio in total consumption expenditures declines, when climate impacts hit only the agricultural sector. This effect increases with the size of the cumulative productivity loss.

The dynamics differ when we focus on the asset-based inequality measures. First, there is only a small increase in the share of wealthless households. The modest increase in wealthless households can be traced back to households that initially held very little wealth, but spend a large share of their consumption expenditures on food. The increase in cost of living due to higher food prices force these households to run down their buffer stock and to increase their consumption expenditures instead to cope with higher food prices. However, only very few households are trapped in these dynamics. This holds for all sizes of productivity losses. Similar dynamics can be observed for the Gini index. Even if rich households increase their asset holdings, households in the middle of the distribution also increase their asset holdings due to higher labor income. As the share of households at the bottom of the distribution does not increase, the total impact on the Gini index is limited. For all loss sizes, the Gini index for wealth remains unchanged, when climate damages solely affect food prices.

In contrast, when climate-related productivity losses are equally distributed across sectors, the agricultural productivity gap, and thus food prices, remain unchanged. In this case, climate impacts lower directly both capital and labor income. Without the hike in food prices inequality decreases measured by the 80-20 total consumption ratio. However, compared to the previous case, the decrease in expenditure-based inequality indicator is driven by changes at the top of the distribution. When climate damages affect incomes directly, rich households decrease their consumption expenditures more drastically than the poorest households. This difference increases in the size of the cumulative productivity loss. Intuitively, the poorest households are close to the subsistence level and thus are less able to decrease their consumption.

With symmetric climate-related productivity losses, the impact on wealth inequality depends on the indicator. For all sizes of loss, the asset-based metrics have opposite signs. While the Gini index for wealth decreases and thus indicates lower wealth inequality, the share of households without any wealth increases in all cases. The share of the population without any assets increases up to 10 percentage points for the highest cumulative loss. When climate damages affect incomes, the asset holdings of all households decline. The decrease in asset holdings is not equally distributed across the distribution. Richer households decrease in absolute terms more their asset holdings. However, as the households with little wealth are affected as well, the decline in the Gini index is modest. Thus, while the Gini index decreases, the impact on households with little wealth is significant, when damages affect only incomes.

Finally, we analyze the case, where climate damages fall entirely on non-agricultural productivity. While the direct impact on incomes is higher than in the symmetric case, households benefit from lower food prices, because climate damages lower the agricultural productivity gap. Similar to the case with symmetric damages, inequality measured by the 80-20 ratio of consumption expenditures decreases. The decrease is driven by the reduction in consumption expenditures of the richest households. Thus, the impact of climate change on incomes dominates the benefits of lower food prices. As the impact on incomes is stronger, when damages fall entirely on non-agricultural productivity, the decrease increases with the size of the cumulative productivity loss. The reduction in consumption inequality for the largest productivity loss is larger than for the other two damage allocations. This supports how income-related damages affect richest households stronger, because the poorest households are close to the subsistence level.

 The impact on wealth inequality differs compared to the expenditure-based inequality metric. When damages affect only non-agricultural productivity, the Gini index for wealth decreases for productivity losses up to 15 percent, but increases for the case with the highest productivity loss. In contrast, the share of households without any wealth increases in the size of the productivity loss. For the lower and medium sized productivity loss, the impact on asset holdings of richer households are more pronounced than for households with little wealth. Consequently, the Gini index decreases in this case. In contrast, for the high loss scenario, the impact on the households with little wealth dominates and the Gini index increases. This is reflected by the increase in the share of households without any wealth. For the highest cumulative loss, the share of households without wealth increases by 15 percentage points. This stresses the importance of income-related damages for wealth inequality.

\newpage

\section{Conclusion}

Climate change poses a significant threat to economic growth and human development. Agricultural productivity losses are not only important for structural change, but may increase poverty through their impact on food prices. However, the long-term distributive impact of climate impacts, is not well understood in this context.
In this paper, we addressed this research gap and analyzed the long-run distributive impact of past climate-related agricultural productivity losses. Climate impacts reduced agricultural productivity and thereby increased food prices in the developing world. Depending on the initial food expenditure share, higher food prices have a direct distributive impact through the partial equilibrium effect. However, food price increases can additionally have an indirect impact on incomes through general equilibrium effects.
General equilibrium effects matter for the distributive impacts of climate-related agricultural productivity losses. The results show that the direct impact of higher food prices effect overestimates the impact for the poorest households. The size of the potential bias in the welfare measurement increases in the size of the impact on agricultural productivity loss. This highlights the importance of considering general equilibrium effects, when analyzing the welfare impact of food price changes. The differences can be traced back to heterogeneous changes in incomes and thus in the level of total consumption expenditures of households. Higher food prices increase the cost of living and thus the precautionary savings motive. For most households the stronger precautionary savings motive dominates the need to increase their total consumption expenditures. Only for the poorest households, who are unable to accumulate assets, the need to increase consumption expenditures dominates in order to cope with the higher cost of basic food consumption. 

Our analysis has important implications for future climate impacts. Depending on the allocation of productivity losses across sectors, different damage channels affect households. Climate impacts may affect households through the impact on food prices, by lowering incomes directly or through both channels simultaneously. The analysis shows how poverty is dynamic and multidimensional. The distributive impact of climate change can differ depending on the inequality indicator used. The results highlight the need for using more than one indicator to capture the dynamic effects of climate damages across the distribution. When climate damages solely affect food prices, the impact on wealth inequality is limited. However, as soon as climate impacts affect incomes, the share of households without any wealth increases in all cases.

The main limitation of the present analysis is related to data availability. Most applications of heterogeneous agent models of the Aiyagari-Bewley-Hugget type focus on advanced economies for specific reasons. Without high quality household panel data that contains detailed information on income sources and wealth holdings, calibrating the model is subject to caveats. However, as the present analysis shows idiosyncratic risk can be an important factor to consider in the context of the developing world. Poverty is a multidimensional phenomenon and heterogeneous agent macroeconomic models can be an important tool to increase the methodological scope. We leave several important aspects for future research. First, climate impacts will affect also other savings motives apart from precautionary saving. Existing literature, indicates that intergenerational aspects such as bequests important in explaining saving behavior of households. Furthermore, the permanent decrease in productivity in Agriculture has also implications for structural change. As capital and labor are perfectly mobile across sectors in our model, agricultural capital and labor shares decrease significantly in our model. However, mobility restrictions could tie labor and capital to the relative unproductive agricultural sector and decrease directly the income of the poorest households and strengthen the adverse impact of the poorest households. Analyzing the implications for structural change in a model with occupational choice and labor market frictions could be promising.

Our analysis indicates how macroeconomic dynamics interact with household-level distributive impact. Bridging the gap between research fields and approaches is essential for improving our understanding of the long-term distributive effects in the developing world. This would be an important step towards achieving the sustainable development goals and improve the livelihood of millions of poor households.

\newpage
\begin{singlespace}
\bibliographystyle{elsarticle-harv}
\bibliography{food_poverty}
\end{singlespace}

\newpage
\appendix

\section{Analytical Derivations}
\label{app_analytical}

\subsection{Household Optimization}
We start by deriving the \textbf{demand functions for both goods (C,F)}. Use the definition of income to get an expression for food:
\[
F= \frac{y}{p} - \frac{C}{p}
\]

Plug it into the utility function and take the derivative with respect to consumption:
\[
U = \frac{(C^\phi(\frac{y}{p} - \frac{C}{p} - \underline{F}))^{1-\phi}}{1 - \eta}
\]
\[
\frac{\partial U}{\partial C} = ( C^\phi(\frac{y}{p} - \frac{C}{p} - \underline{F}^{1-\phi}))^{-\phi}[\phi C^{\phi-1}(\frac{y}{p} - \frac{C}{p} - \underline{F})^{1-\phi} + C^\phi (1-\phi) (\frac{y}{p} - \frac{C}{p} - \underline{F})^{-\phi}(-\frac{1}{p})] = 0
\]
Now, rearrange and solve for C:

\[
\phi C^{\phi-1}(\frac{y}{p} - \frac{C}{p} - \underline{F})^{1-\phi} = \frac{C^\phi}{p}  (1-\phi) (\frac{y}{p} - \frac{C}{p} - \underline{F})^{-\phi} 
\]

\[
\frac{\phi}{C}(\frac{y}{p} - \frac{C}{p} - \underline{F}) = \frac{1 - \phi}{p}
\]
\[
C (\frac{1}{1-\phi}) = \frac{\phi}{1-\phi} (y - p \underline{F})
\]
\[
C = \phi (y - p \underline{F})
\]
Combining the demand function for consumption and the definition of income we get the demand function for food:
\[
F= \frac{y}{p} - \frac{\phi (y - p \underline{F})}{p} = \frac{(1-\phi)y}{p} + \phi \underline{F}
\]

Next, we can plug both demand functions into the utility function and thus express it in terms of the price and income. The \textbf{indirect utility function} will become useful in the solution of the model and can be derived as follows:
\[
U(p,y) = \frac{(( \phi (y - p \underline{F}))^\phi(\frac{(1-\phi)y}{p} + \phi \underline{F} -\underline{F})^{1-\phi}))^{1 - \eta}}{1 - \eta}
\]

\[
U(p,y) = \frac{1}{1 - \eta}\bigg(( \phi (y - p \underline{F}))^\phi(\frac{(1-\phi)y}{p} -(1-\phi)  \underline{F})^{1-\phi})\bigg)^{1 - \eta}
\]

\[
U(p,y) = \frac{1}{1 - \eta}\bigg(( \phi (y - p \underline{F}))^\phi(\frac{(1-\phi)}{p}(y -p\underline{F}))^{1-\phi}\bigg)^{1 - \eta}
\]

\[
U(p,y) = \frac{1}{1 - \eta}\bigg(( \phi^\phi (y - p \underline{F})^\phi (1-\phi)^{1-\phi} p^{\phi-1} (y -p\underline{F})^{1-\phi})\bigg)^{1 - \eta}
\]

Define $ \Phi =\phi^\phi (1-\phi)^{1-\phi} $ and simplify to get:

\[
U(p,y) = \frac{1}{1 - \eta}\bigg( \Phi  p^{\phi-1} (y -p\underline{F})\bigg)^{1 - \eta}
\]

We can write the optimization problem of households recursively with the following Bellman equation:

\[
V(a,\theta)= \max_{a'\geq 0} \frac{1}{1 - \eta} \bigg( \Phi  p^{\phi-1}\bigg)^{1 - \eta} (r a + \theta w - a' -p\underline{F})^{1 - \eta} + \beta E_{\theta'} V(a',\theta')
\]

The first order condition is given by:

\[
\frac{\partial V}{\partial a'} =  \bigg( \Phi  p^{\phi-1}\bigg)^{1 - \eta} (r a + \theta w - a' -p\underline{F})^{ - \eta} (-1) + \beta \frac{\partial V(a',\theta')}{\partial a'} = 0
\]

\[
 \bigg( \Phi  p^{\phi-1}\bigg)^{1 - \eta} (r a + \theta w - a' -p\underline{F})^{ - \eta} = \beta \frac{\partial V(a',\theta')}{\partial a'}
\]

We want to substitute the term $\frac{\partial V(a',\theta')}{\partial a'}$, because it is a derivative for a yet unknown function. For that we can derive the Envelope Condition.

\[
\frac{\partial V(a,\theta)}{\partial a}= \bigg( \Phi  p^{\phi-1}\bigg)^{1 - \eta} (r a + \theta w - a' -p\underline{F})^{ - \eta} R = \bigg( \Phi  p^{\phi-1}\bigg)^{1 - \eta} (y(a,\theta) -p\underline{F})^{ - \eta}R
\]
Update one period to get:

\[
\frac{\partial V(a',\theta')}{\partial a'}=  \bigg( \Phi  p^{\phi-1}\bigg)^{1 - \eta} (y(a',\theta') -p\underline{F})^{ - \eta}R
\]

Plug it into the FOC to get:
\[
 \bigg( \Phi  p^{\phi-1}\bigg)^{1 - \eta} (r a + \theta w - a' -p\underline{F})^{ - \eta} = \beta \bigg( \Phi  p^{\phi-1}\bigg)^{1 - \eta} (y(a',\theta') -p\underline{F})^{ - \eta}R
\]
Plug it into the FOC to get:
\[
 (y(a,\theta) -p\underline{F})^{ - \eta} = \beta  (y(a',\theta') -p\underline{F})^{ - \eta}R
\]
where 
\[
y(a,\theta)= r a + w \theta - a'
\]
and 
\[
y(a',\theta')= y(r a + w \theta - y(a,\theta),\theta')
\]
\subsection{Consumption Equivalent Variation}

The equivalent satisfies the following condition:\\

Indirect Utility:
\[
I(p_0,y_0)=I(p,y+EV)
\]
\[
\frac{1}{1 - \eta}\bigg( \Phi  p_0^{\phi-1} (y_0 -p_0\underline{F})\bigg)^{1 - \eta}=\frac{1}{1 - \eta}\bigg( \Phi  p^{\phi-1} (y+EV -p\underline{F})\bigg)^{1 - \eta}
\]
\[
 p_0^{\phi-1} (y_0 -p_0\underline{F})=  p^{\phi-1} (y+EV -p\underline{F})
\]
\[
EV= (\frac{p_{0}}{p})^{\phi-1} (y_0 -p_0\underline{F})-(y -p\underline{F})
\]
In Partial Equilibrium $y=y_0$, thus:
\[
EV_{PE}=(\frac{p_{0}}{p})^{\phi-1} (y_0 -p_0\underline{F})-(y_0 -p\underline{F})
\]

\[
EV_{PE}= (\frac{p_{0}}{p})^{\phi-1} y_0 -\frac{1}{p})^{\phi-1}p_0^\phi \underline{F}-y_0 -p\underline{F}
\]
\[
EV_{PE}= ((\frac{p_{0}}{p})^{\phi-1}-1) y_0   +(p -{p}^{1-\phi}p_0^\phi)\underline{F}
\]
\section{Comparison Productivity loss}
\label{app_Comp}
\begin{table}[h!]

\begin{tabularx}{\textwidth}{bsssssss}
\hline \hline
Scenario & $f^{8020}$ & $\mu^f$ & $y^{8020}$  &  $gini_W$ &  $\Delta Y^f$ &  $\Delta Y^c$ &  $\Delta K$\\
\hline
Baseline & $-1.11$  & $0.0311$ & $-3.14$ & $0.00007$ & $-0.1303$ & $-0.015$ & $0.028$ \\
Low Loss & $-0.47$  & $0.0142$ & $-1.4$ & $0.0003$ & $-0.06$ & $0.001$ & $0.0265$ \\
High Loss & $-1.77$  & $0.0466$ & $-4.83$ & $0.0009$ & $-0.1955$ & $-0.003$ & $0.0273$ \\

    \hline
  \end{tabularx}

\caption{Comparison of baseline results ($\xi^f=0.25$) to low ($\xi^f = 0.11$) and high estimates ($\xi^f = 0.4$) for key aggregate variables: 80-20 ratio of food expenditures, mean share of food expenditures, 80-20 ratio of total consumption expenditures, Gini index wealth, agricultural output (percentage change), non-agriculture output (percentage change), aggregate capital (percentage change)}
\end{table}
\section{Additional Figures}
\label{app_fig}

\begin{figure}
    \centering
    \includegraphics[scale =0.6]{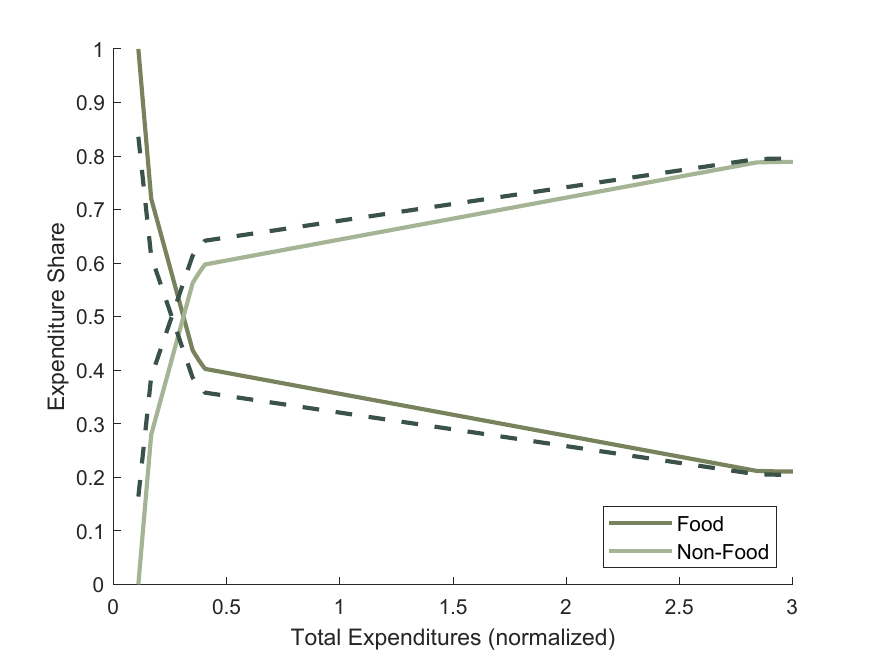}
    \caption{Changes in consumption pattern across the distribution for the partial equilibrium outcome}
\label{det_micro_transfers}
\end{figure}

\begin{figure}[h!]
    \centering
    \includegraphics[scale=0.75]{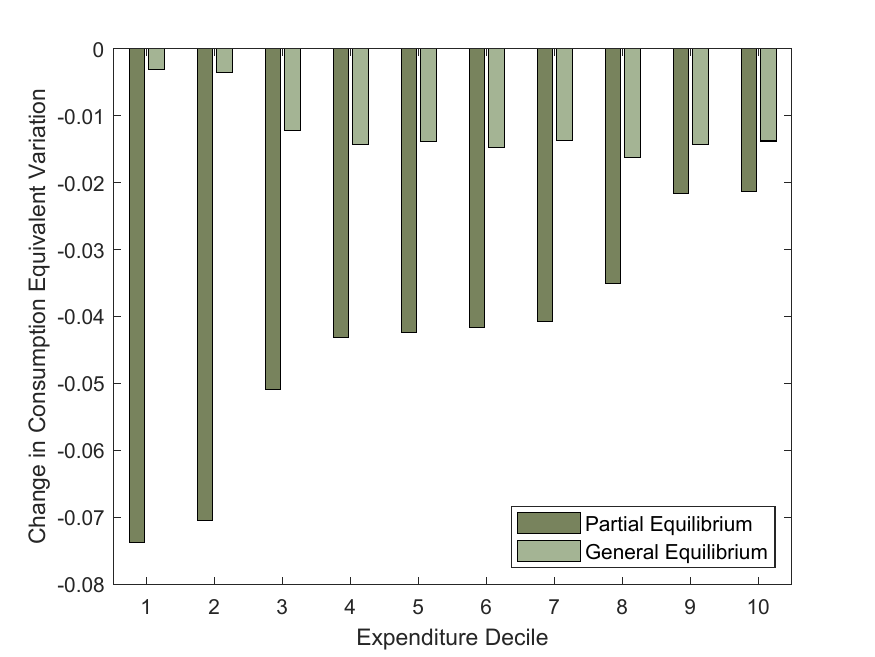}
    \caption{Difference in welfare outcome between partial and general equilibrium outcome for the low estimate ($\xi^f=0.11$). A negative value indicates that the welfare loss in partial equilibrium is higher than in general equilibrium}
    \label{DiffPEGE_low}
\end{figure}

\begin{figure}[h!]
    \centering
    \includegraphics[scale=0.75]{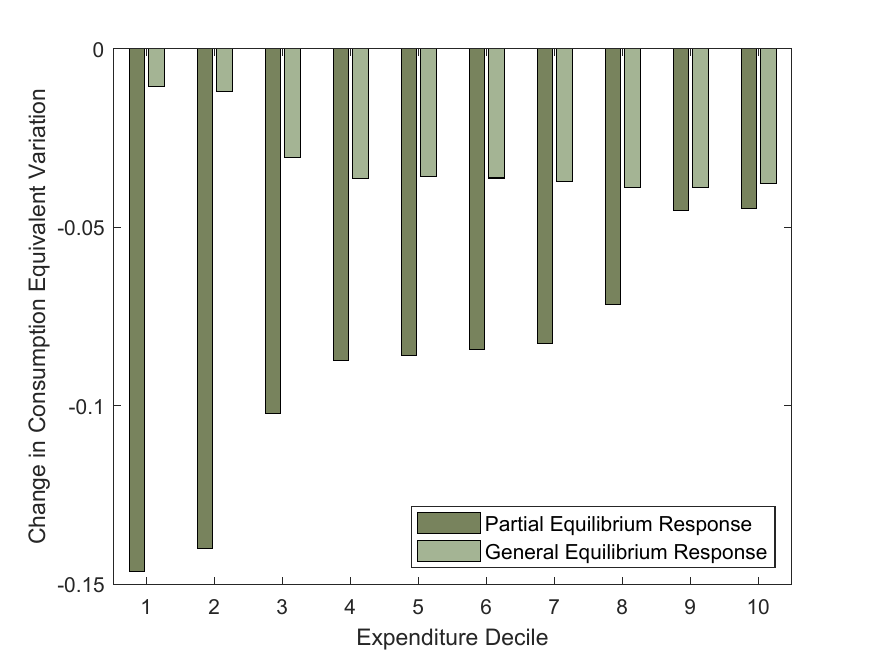}
    \caption{Difference in welfare outcome between partial and general equilibrium outcome for the high estimate ($\xi^f=0.4$). A negative value indicates that the welfare loss in partial equilibrium is higher than in general equilibrium}
    \label{DiffPEGE_high}
\end{figure}

\end{document}